\documentclass[fleqn,usenatbib]{mnras}

\usepackage{newtxtext,newtxmath}

\usepackage[T1]{fontenc}
\usepackage{ae,aecompl}

\usepackage{graphicx}	
\usepackage{amsmath}	
\usepackage{amssymb}	

\newcommand{\orcid}[1]{\href{https://orcid.org/#1}{\textcolor[HTML]{A6CE39}{\aiOrcid}}}

\title[$u$-band catalogue of the $AKARI$ NEP-Wide]{CFHT MegaPrime/MegaCam $u$-band source catalogue of the $AKARI$ North Ecliptic Pole Wide field}

\author[T.-C. Huang et al.]{
Ting-Chi Huang \href{https://orcid.org/0000-0001-7200-8157}{\includegraphics[scale=0.1]{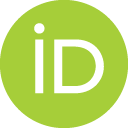}},$^{1,2}$\thanks{E-mail: kevintch@ir.isas.jaxa.jp}
Hideo Matsuhara,$^{1,2}$
Tomotsugu Goto \href{https://orcid.org/0000-0002-6821-8669}{\includegraphics[scale=0.1]{ORCIDiD_icon128x128.png}},$^{3}$
Hyunjin Shim,$^{4}$
\newauthor
Seong Jin Kim \href{https://orcid.org/0000-0001-9970-8145}{\includegraphics[scale=0.1]{ORCIDiD_icon128x128.png}},$^{3}$
Matthew A. Malkan \href{https://orcid.org/0000-0001-6919-1237}{\includegraphics[scale=0.1]{ORCIDiD_icon128x128.png}},$^{5}$
Tetsuya Hashimoto \href{https://orcid.org/0000-0001-7228-1428}{\includegraphics[scale=0.1]{ORCIDiD_icon128x128.png}},$^{3,6}$
\newauthor
Ho Seong Hwang \href{https://orcid.org/0000-0003-3428-7612}{\includegraphics[scale=0.1]{ORCIDiD_icon128x128.png}},$^{7}$
Nagisa Oi \href{https://orcid.org/0000-0002-4686-4985}{\includegraphics[scale=0.1]{ORCIDiD_icon128x128.png}},$^{8}$
Yoshiki Toba \href{https://orcid.org/0000-0002-3531-7863}{\includegraphics[scale=0.1]{ORCIDiD_icon128x128.png}},$^{9,10,11}$
Dongseob Lee,$^{4}$
\newauthor
Daryl Joe D. Santos \href{https://orcid.org/0000-0002-5687-0609}{\includegraphics[scale=0.1]{ORCIDiD_icon128x128.png}},$^{3}$
and Toshinobu Takagi$^{2,12}$
\\
$^{1}$Department of Space and Astronautical Science, Graduate University for Advanced Studies, SOKENDAI, Shonankokusaimura, \\
Hayama, Miura District, Kanagawa 240-0193, Japan\\
$^{2}$Institute of Space and Astronautical Science, Japan Aerospace Exploration Agency, 3-1-1 Yoshinodai, Chuo-ku, Sagamihara, \\
Kanagawa 252-5210, Japan\\
$^{3}$Institute of Astronomy, National Tsing Hua University, No. 101, Section 2, Kuang-Fu Road, Hsinchu City 30013, Taiwan\\
$^{4}$Department of Earth Science Education, Kyungpook National University, 80 Daehak-ro, Buk-gu, Daegu 41566, Republic of Korea\\
$^{5}$Department of Physics and Astronomy, UCLA, 475 Portola Plaza, Los Angeles, CA 90095-1547, USA\\
$^{6}$ Centre for Informatics and Computation in Astronomy (CICA), National Tsing Hua University, 101, Section 2. Kuang-Fu Road, \\
Hsinchu City 30013, Taiwan\\
$^{7}$Korea Astronomy and Space Institute, 776 Daedeok-daero, Yuseong-gu, Daejeon 34055, Republic of Korea \\
$^{8}$Tokyo University of Science, 1-3 Kagurazaka, Shinjuku-ku, Tokyo 162-8601, Japan\\
$^{9}$Department of Astronomy, Kyoto University, Kitashirakawa-Oiwake-cho, Sakyo-ku, Kyoto 606-8502, Japan\\
$^{10}$Academia Sinica Institute of Astronomy and Astrophysics, 11F of Astronomy-Mathematics Building, AS/NTU, No. 1, Section 4, \\
Roosevelt Road, Taipei 10617, Taiwan\\
$^{11}$Research Center for Space and Cosmic Evolution, Ehime University, 2-5 Bunkyo-cho, Matsuyama, Ehime 790-8577, Japan\\
$^{12}$Japan Space Forum, 3-2-1, Kandasurugadai, Chiyoda-ku, Tokyo 101-0062, Japan\\
}

\date{Accepted XXX. Received YYY; in original form ZZZ}

\pubyear{2020}

\begin{document}
\label{firstpage}
\pagerange{\pageref{firstpage}--\pageref{lastpage}}
\maketitle

\begin{abstract}
The $AKARI$ infrared (IR) space telescope conducted two surveys (Deep and Wide) in the North Ecliptic Pole (NEP) field to find more than 100,000 IR sources using its Infrared Camera (IRC). IRC's 9 filters, which cover wavebands from 2 to 24 $\mu$m continuously, make $AKARI$ unique in comparison with other IR observatories such as $Spitzer$ or $WISE$. However, studies of the $AKARI$ NEP-Wide field sources had been limited due to the lack of follow-up observations in the ultraviolet (UV) and optical. In this work, we present the Canada-France-Hawaii Telescope (CFHT) MegaPrime/MegaCam $u$-band source catalogue of the $AKARI$ NEP-Wide field. The observations were taken in 7 nights in 2015 and 2016, resulting in 82 observed frames covering 3.6 deg$^2$. The data reduction, image processing and source extraction were performed in a standard procedure using the \textsc{Elixir} pipeline and the \textsc{AstrOmatic} software, and eventually 351,635 sources have been extracted. The data quality is discussed in two regions (shallow and deep) separately, due to the difference in the total integration time (4,520 and 13,910 seconds). The 5$\sigma$ limiting magnitude, seeing FWHM, and the magnitude at 50 per cent completeness are 25.38 mag (25.79 mag in the deep region), 0.82 arcsec (0.94 arcsec) and 25.06 mag (25.45 mag), respectively. The u-band data provide us with critical improvements to photometric redshifts and UV estimates of the precious infrared sources from the $AKARI$ space telescope. 
\end{abstract}




\section{Introduction}
The study of galaxy evolution encompasses some of the major unsolved problems in modern astronomy. For example, the peak and the overall characteristics of the cosmic star formation rate history still remain uncertain, partly because of the effects of interstellar dust \citep[e.g.][]{Madau2014}. In order to improve the measurements from previous studies of galaxy evolution, having a large multi-wavelength galaxy sample ranging from low- to high-redshift is essential.

Star formation (SF) and active galactic nuclei (AGN) play an important role in galaxy evolution. Both of them tend to be obscured by dust, and thus produce powerful infrared emissions. To uncover the SF and AGN activities hidden by dust, the $AKARI$ infrared astronomy satellite, developed by the Japan Aerospace Exploration Agency (JAXA), was launched in 2006 \citep{Murakami2007}. $AKARI$ carried out a survey in the NEP field \citep{Matsuhara2006} using its IRC \citep{Onaka2007}. $AKARI$ IRC's 9 filters: $N2$, $N3$, $N4$, $S7$, $S9W$, $S11$, $L15$, $L18$, and $L24$, continuously cover the observed wavelengths from 2 to 24 $\mu$m, which is a very crucial range to identify key mid-IR features of SF or AGN, such as polycyclic aromatic hydrocarbon emissions, the 9.7-$\mu$m silicate absorption, and the AGN hot dust emission \citep[e.g.][]{Smith2007,Lacy2020}. In the sense of the observation range, $AKARI$ surpasses other IR space telescopes, for example, $Spizter$ and $WISE$, because both of them have an observation gap at this important wavelength. There were two major observing strategies of the $AKARI$ NEP survey, a deep survey (hereafter NEP-Deep) in a 0.5-deg$^2$ area \citep{Wada2008,Takagi2012,Murata2013} and a wide survey (hereafter NEP-Wide) in a 5.4-deg$^{2}$ area \citep{Lee2009,Kim2012}. 

Multi-wavelength follow-up observations have been performed in the NEP-Deep from facilities and instruments such as CFHT MegaCam \citep[][hereafter O14]{Oi2014} and Chandra ACIS-I \citep{Krumpe2015}. Many valuable studies have been conducted, such as the determination of IR luminosity function and star formation history \citep{Goto2015}, AGN selection by fitting IR spectral energy distributions \citep{Huang2017}, and clustering of extremely red objects \citep{Seo2019}. However, in contrast to the extensively surveyed NEP-Deep, the NEP-Wide did not have an uniform optical or near-UV observation. It had been only observed by MegaCam in a relatively small area of 2 deg$^{2}$ at the center \citep[][hereafter, H07]{Hwang2007}, despite that the outer area had been compensated by shallower observations from Maidanak Observatory \citep{Jeon2010}. Consequently, it had suffered from lack of deep optical and near-UV photometry for several years. It was not until deep Subaru Hyper Suprime-Cam \citep[HSC;][]{Miyazaki2018} observations were carried out \citep[PI: T. Goto;][]{Goto2017} in 2014 and 2015, that we obtained the deep optical ancillary data over the entire AKARI NEP-Wide field (Oi et al. 2020).

\begin{figure}
    \centering
    \includegraphics[width=\columnwidth]{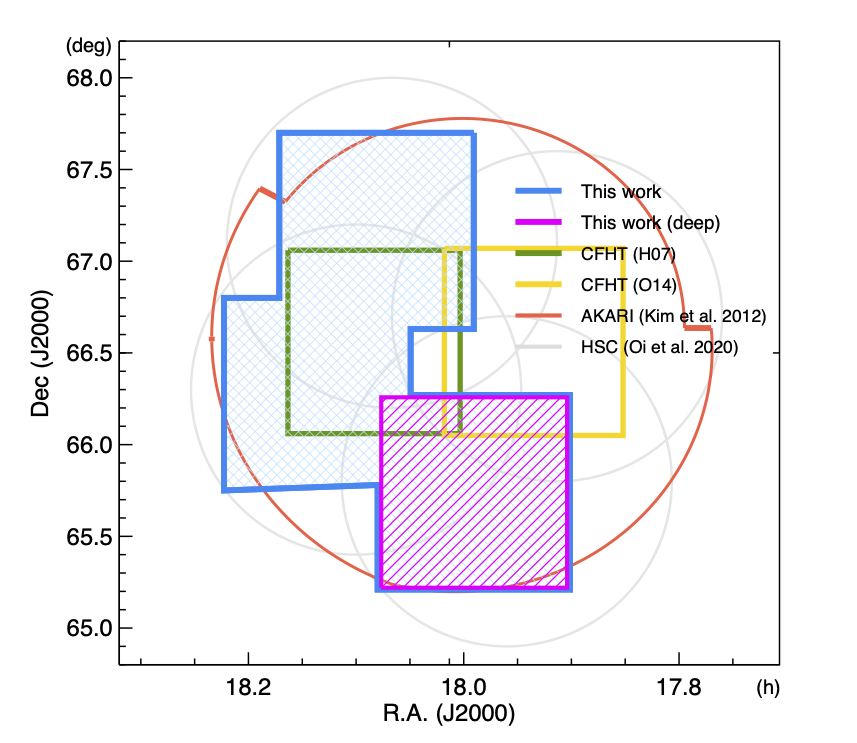}
    \caption{The coverage of the surveys related to this work in the $AKARI$ NEP field (red), including previous $u^{*}$-band observations (green and yellow), HSC observation (grey) and this work (blue and purple).}
    \label{fig:obs}
\end{figure}

Aside from the HSC's 5 optical bands ($g$, $r$, $i$, $z$, $Y$), having an additional band in the near-UV is very crucial for further research, because it helps the identification of the Balmer break in the photometric redshift (photo-$z$) calculation. \cite{Sawicki2019} showed that including the CFHT $u$-band into the HSC 5-band sample significantly improves the photo-$z$ performance. However, as mentioned previously, the only available near-UV data in the $AKARI$ NEP field were the $u^{*}$-band data from MegaCam, but they were limited to small observed areas, which only cover approximately 1 deg$^{2}$ (H07) and 0.67 deg$^{2}$ \citep[][O14]{Takagi2012}. Therefore, a new CFHT MegaPrime/MegaCam $u$-band observation on the $AKARI$ NEP-Wide was proposed, and was completed successfully in 2016. Fig.~\ref{fig:obs} shows the coverage of the previous $u^{*}$-band observations, the HSC observation, and the latest $u$-band observation (this work) we obtained in the $AKARI$ NEP field.

The $u$-band data from this work is being used in a variety of studies. They are included in the band-merged catalogue of the $AKARI$ NEP field (Kim et al. 2020 submitted), and furthermore, with the multi-wavelength photometry, photo-$z$'s have been calculated using \textsc{LePHARE} (Ho et al. 2020 submitted). Additionally, research in the $AKARI$ NEP field is also pursuing further topics, for example, the search for optically dark IR galaxies \citep{Toba2020}, the selection of IR AGN (Wang et al. 2020 submitted), galaxy clusters (Huang et al. in preparation), and environmental effects on AGN activity (Santos et al. in preparation). 

This paper is organised as follows. The observations and data reduction are described in Sect.~\ref{method}. We present the limiting magnitude, seeing, completeness, and source number counts in Sect.~\ref{result}. We then compare results with the previous $u^{*}$-band surveys (i.e. H07 and O14) in the AKARI NEP field in Sect.~\ref{dis}. A summary is given in Sect.~\ref{sum}. Throughout this paper, we use the AB magnitude system and the J2000.0 equatorial coordinate system. The letters $u^{*}$ and $u$ refer to the CFHT MegaCam old filter and the MegaPrime/MegaCam new filter, respectively.

\section{Method}
\label{method}
\subsection{Observation}
\label{obs}
The observations of the $AKARI$ NEP field were made using the CFHT MegaPrime/MegaCam with its $u$-band filter (u.MP9302; range 3041-4014\AA{}; effective wavelength at 3610\AA{}) in the CFHT Queued Service Observations mode. The MegaPrime/MegaCam detector has 40 CCDs, and each CCD is composed of 2048 $\times$ 4612 pixels with a pixel scale of 0.185 arcsec. There were two observing runs. One was from 2015 May 22nd to 26th and the other was from 2016 July 6th and 7th (PI:T. Goto). The total observed area was about 3.6 square degrees. The former observation run contained 6 observation pointings--NEP1, NEP2, NEP3, NEP4, NEP5, and NEP6, while the latter had three--NEP1A, NEP1B, and NEP2A. The details of each observations are summarised in Table.~\ref{tab:obs}, including coordinates, integration time per frame, number of frames, ambient temperature, and humidity. The coordinates and the integration time are the average value of all frames, while the ambient temperature and the humidity refer to the first observed frame in each pointing. The NEP1, NEP2, NEP1A, NEP1B, and NEP2A pointings targeted the same region, providing a total integration time of 13,910 seconds. The NEP3 and NEP4, and the NEP5 and NEP6 pointings both have a 4,520-second total integration time. 

\begin{table*}
	\centering
	\caption{The summary of the observation pointings. The columns from the left to right are name, right ascension, declination, integration time, number of frames, ambient temperature, and humidity.}
	\label{tab:obs}
	\begin{tabular}{lcccccr} 
		\hline
		Name &  $\alpha$ (deg)  & $\delta$ (deg) & Int. Time (sec) & $\#$ of Frames & Temp. ($^{\circ}$C) & Humidity ($\%$) \\
		\hline
		NEP1 & 269.72607 & +65.691054 & 300.2 & 13 & 1.6 & 11.6\\
		NEP1A & 269.72560 & +65.699449 & 300.2 & 9 & 5.6 & 28.5\\
		NEP1B & 269.72798 & +65.766516 & 300.2 & 10 & 5.8 & 8.7\\
		NEP2 & 269.97564 & +65.756393 & 265.2 & 8 & 4.0 & 27.5\\
		NEP2A & 269.97386 & +65.758351 & 219.2 & 10 & 6.4 & 7.5\\
		NEP3 & 271.26471 & +67.223669 & 300.2 & 8 & 4.7 & 21.1\\
		NEP4 & 271.01165 & +67.164424 & 265.2 & 8 & 4.1 & 38.6\\
		NEP5 & 272.07713 & +66.273413 & 300.2 & 8 & 3.4 & 42.7\\
		NEP6 & 271.81885 & +66.331610 & 265.2 & 8 & 2.0 & 46.3\\
		\hline
	\end{tabular}
\end{table*}

\subsection{Data Reduction}
The data reduction was performed by the \textsc{Elixir} pipeline \citep{Magnier2004} in a regular procedure, which includes the bad pixel mask, the bias structure removal, the flat-field correction, and the overscan region elimination. 

The co-added image was obtained by using the \textsc{AstrOmatic} software, including \textsc{WeightWatcher} \citep{Marmo2008}, \textsc{SExtractor} \citep{Bertin1996}, \textsc{Scamp} \citep{Bertin2006}, and \textsc{SWarp} \citep{Bertin2002}. The overall procedure was that we first applied \textsc{WeightWatcher} to every processed image to create the weight maps for them. Then, we used \textsc{SExtractor} to extract sources from every image and listed them into the catalogues in the `fits$\_$ldac' format for \textsc{Scamp} to read in the next step. We used \textsc{Scamp} to calibrate the astrometry and the magnitude zero-point based on the input `fits$\_$ldac' catalogues. Finally, \textsc{SWarp} mosaicked images from dithering into a final image of the whole observed field. 

After \textsc{SWarp} constructed the final co-added image, we ran \textsc{SExtractor} again on the co-added image as the input, to obtain the final source catalogue. Finally, we calibrated the magnitude zero-point with the data from Sloan Digital Sky Survey Data Release 16 \citep[SDSS DR16;][]{Ahumada2020}. The details of each procedure are described in the following sections.

\subsubsection{Weight map: WeightWatcher}
\label{ww}
\begin{figure}
    \centering
    \includegraphics[width=\columnwidth]{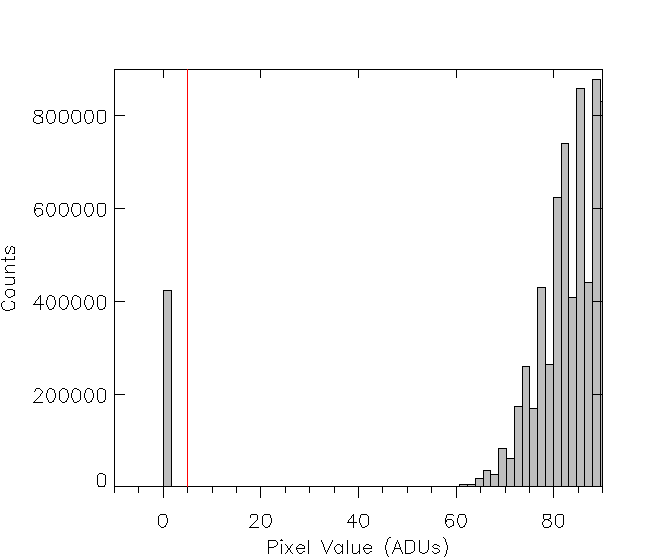}
    \caption{The number distribution of the pixel values in one CCD of one observed frame at the faint end. The red vertical line indicates the pixel value of 5 ADUs.}
    \label{fig:pix}
\end{figure}

\textsc{WeightWatcher} generates a weight map for an input image, that is, it gives every pixel a weight value (0 or 1), so that the defects on pixels can be avoided. For creating the weight maps, the lower and upper thresholds to give a 0 value can be set by the configuration parameters \texttt{WEIGHT$\_$MIN} and \texttt{WEIGHT$\_$MAX}. The pixel values of our processed images are generally greater than 50, so we defined the lower threshold to be 5 ADUs (i.e. \texttt{WEIGHT$\_$MIN}=5). Fig.~\ref{fig:pix} shows the distribution of the pixel values in one CCD chip of a single exposure frame at the faint end, as an example to justify our lower threshold choice. On the other hand, we did not give the 0 value to bright pixels, so the upper threshold was equivalent to the saturation value, 65535 ADUs.

\subsubsection{Astrometric $\&$ photometric calibration: Scamp}
\label{scamp}
\textsc{Scamp} calibrates the astrometry and zero-point magnitude. For astrometric calibration, we used the default CDS server (cocat1.u-strasbg.fr) for the reference catalogue. We referred to the Guide Star Catalogue version 2.3 \citep{Lasker2008}, and the nearest band, $U$ band (range 300-400 nm), was used. Since the instrument contains multiple CCD chips, we had to input the world coordinate system information using the additional header files `*.ahead' for every frames. We first ran the \textsc{Scamp} with the mosaic type LOOSE to generate the `ahead' files and then set the mosaic type to SAME$\_$CRVAL for a second run to obtain the astrometric solution.

For photometric calibration, the magnitude $m$ is calculated as
\begin{equation}
    m = -2.5\textrm{log}(f) + m_{0} + 2.5 \textrm{log}(t) + c \times a,
\end{equation}
where $f$ is the measured flux (in ADUs), $m_{0}$ is the instrumental magnitude zero-point for a one-second exposure (in mag), $t$ is the exposure time (in sec), $c$ is the extinction coefficient, and $a$ is the averaged airmass. The input variables are referred from the FITS header keywords of the images and the `fits$\_$ldac' catalogues. For the magnitude output flux scaling, we set the arbitrary magnitude zero-point (\texttt{MAGZERO$\_$OUT}) to be 25.188 mag, which is the same value as the FITS header keyword \texttt{PHOT$\_$C}, the instrumental zero-point measured by \textsc{Elixir}, for a one-second exposure.

\subsubsection{Co-added image: SWarp}
\textsc{SWarp} resamples the input images and stacks them into a co-added image. The size of the output pixels was set to be the median of the input pixel scales at the center of frames (\texttt{PIXELSCALE$\_$TYPE}=MEDIAN). We adopted the resampling method LANCZOS3, and used the median value for the image mosaicking. The oversampling was set to be used  automatically. We allowed interpolation where there is any bad input pixel. Background was subtracted using a 128-pixel by 128-pixel mesh size. The weight maps generated by \textsc{WeightWatcher} in Section \ref{ww} were included in the co-addition process using inverse variance weighting (\texttt{WEIGHT$\_$TYPE}=MAP$\_$WEIGHT). SWarp also generated an inverse variance weight map of the co-added image, which was used as an input in the next source extraction process. 

\subsubsection{Source extraction: SExtractor}
\label{se}
\textsc{SExtractor} extracts sources from an image and performs photometry of the sources. Since we intend to use the $u$-band data together with the optical counterparts observed by HSC (Oi et al. 2020 submitted), which is constructed from deeper imaging, we chose to use a filter with a narrow Gaussian FWHM and a small convolution mask, in order not to miss any possible faint sources. We had compared the source extraction results using several filter files, and eventually we adopted a Gaussian PSF with 2-pixel FWHM and a 3-pixel by 3-pixel convolution mask (Gauss2.0$\_$3x3). The detailed comparison of convolution filter usage is described in Appendix \ref{apdx}. Other parameters related to the source extraction are listed in Table~\ref{tab:se}. For photometry, we adopted the Kron photometry for the source catalogue. The Kron flux (FLUX$\_$AUTO) is defined as the sum of pixel values inside the Kron elliptical aperture $K$, 
\begin{equation}
    \textrm{FLUX\_AUTO} = \sum_{\textrm{i}\in K}^{} p_{\textrm{i}},
\end{equation}
where $p$ is the pixel value in the input image.
The conversion between the Kron flux and the Kron magnitude (MAG\_AUTO) is
\begin{equation}
\label{eq:mag}
\begin{split}
    \textrm{MAG\_AUTO} & = \textrm{MAG\_ZEROPOINT} \\
    & - 2.5\times \textrm{log}(\textrm{FLUX\_AUTO}),
\end{split}
\end{equation}
where the \texttt{MAG$\_$ZEROPOINT} is the input parameter for the magnitude zero-point setting in \textsc{SExtractor}. We used a zero-point value 25.188 as the \textsc{Scamp}'s output zero-point. The saturation level (\texttt{SATUR$\_$LEVEL}) was set to be 50000 ADUs. 

We additionally provide a catalogue based on the coordinates of HSC sources (hereafter, HSC-based source catalogue), using the \texttt{ASSOC$\_$*} parameters in \textsc{SExtractor}. The limiting magnitudes of the HSC 5 bands ($g$, $r$, $i$, $z$, and $Y$) are 28.6, 27.3, 26.7, 26.0, and 25.6, respectively (Oi et al. 2020 submitted). The HSC sources in this analysis are required to have 5$\sigma$ detections in all 5 bands. We adopted the ASSOC process to match sources with the nearest ones (\texttt{ASSOC$\_$TYPE}=NEAREST), and the matching radius (\texttt{ASSOC$\_$RADIUS}) was set to be 6 pixels (i.e. 1.11 arcsec). As a result, this HSC-based catalogue contains 351,635 sources, and among them there are 259,335 matches. The details of the $u$-band and the HSC-based source catalogues are presented in Appendix~\ref{cat}.

\begin{table}
	\centering
	\caption{The extraction parameters in the input configuration file for \textsc{SExtractor}.}
	\label{tab:se}
	\begin{tabular}{lr} 
		\hline
		Parameter & Value \\
		\hline
		\texttt{DETECT$\_$TYPE} & CCD \\
		\texttt{DETECT$\_$MINAREA} & 5 \\
		\texttt{DETECT$\_$MAXAREA} & 0 \\
		\texttt{DETECT$\_$THRESH} & 1.5 \\
		\texttt{ANALYSIS$\_$THRESH} & 1.5 \\
		\texttt{THRESH$\_$TYPE} & RELATIVE \\
		\texttt{FILTER} & Y \\
		\texttt{FILTER$\_$NAME} & gauss$\_$2.0$\_$3x3.conv \\
		\texttt{FILTER$\_$THRESH} & 1.0 \\
		\texttt{DEBLEND$\_$NTHRESH} & 32 \\
		\texttt{DEBLEND$\_$MINCONT} & 0.001 \\
		\texttt{CLEAN} & Y \\
		\texttt{CLEAN$\_$PARAM} & 1.0 \\
		\texttt{MASK$\_$TYPE} & CORRECT \\
		\hline
	\end{tabular}
\end{table}

\subsubsection{Zero-point calibration}
\label{zpcorr}
\begin{figure*}
	\includegraphics[width=\columnwidth]{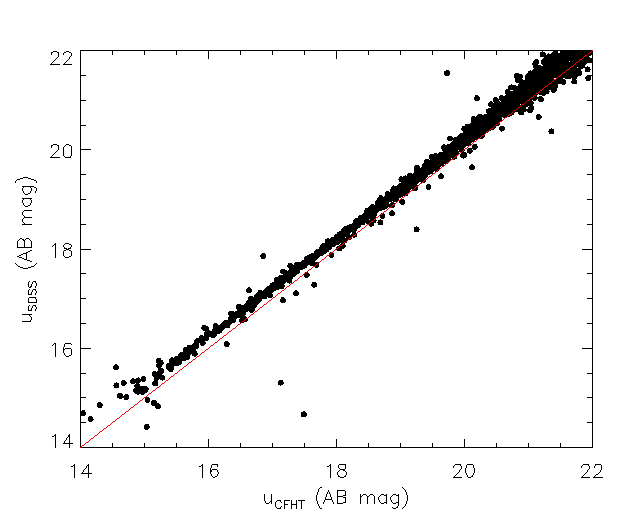}
	\includegraphics[width=\columnwidth]{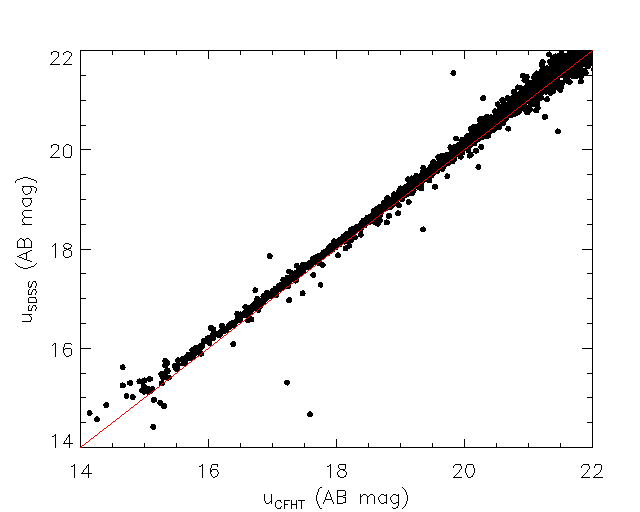}
    \caption{The magnitude comparisons of the red stars between the CFHT $u$ band (this work) and the SDSS $u$ band (DR16). The left and right panels show the magnitude comparisons before and after the zero-point correction, respectively. In each panel, a line with a slope of one is plotted in red.}
    \label{fig:zpcorr}
\end{figure*}

After establishing a source catalogue, it is generally beneficial to perform a magnitude zero-point calibration with an additional catalogue when available. Calibrating with the SDSS is the best choice for CFHT MegaCam photometry, as many previous CFHT studies suggested \citep[e.g.][]{Gwyn2012,Ibata2017,Sawicki2019}. Fortunately, approximately 35,000 sources in the southern part ($\delta$ < +66:45:00) of the AKARI NEP field are observed in the latest SDSS DR16. Among them, 14,568 sources match with our CFHT $u$-band data within a 0.6-arcsec radius. 2,634 sources which have both CFHT and SDSS $u$-band magnitudes less than 22 are selected for the calibration. We further selected 1,607 red stars using $u_{\textrm{SDSS}} - g_{\textrm{SDSS}} > 1.2$ and CLASS$\_$STAR $> 0.95$. The CFHT $u$-band magnitude of the red stars were calibrated with the SDSS $u$-band and $g$-band magnitudes, using the following equation \citep{Sawicki2019}:
\begin{equation}
    u_{\textrm{CFHT}} = u_{\textrm{SDSS}} + 0.036 \times (u_{\textrm{SDSS}}-g_{\textrm{SDSS}}) - 0.165.
\end{equation}
The mean offset of the red stars derived from the equation is about $-0.100$ (i.e. $u_{\textrm{CFHT}} = u_{\textrm{SDSS}} - 0.100$). We performed the zero-point calibration by adding this 0.100 value to our $u$-band magnitude. Fig.~\ref{fig:zpcorr} shows the comparisons between the CFHT u-band magnitude and the SDSS u-band magnitude without/with (left/right) the zero-point calibration. In this calibration, we did not take account of the effects of Galactic extinction. We believe this approximation is allowable, because the latitude of the AKARI NEP field is high.

\section{Result}
\label{result}
\begin{figure*}
	\includegraphics[width=\columnwidth]{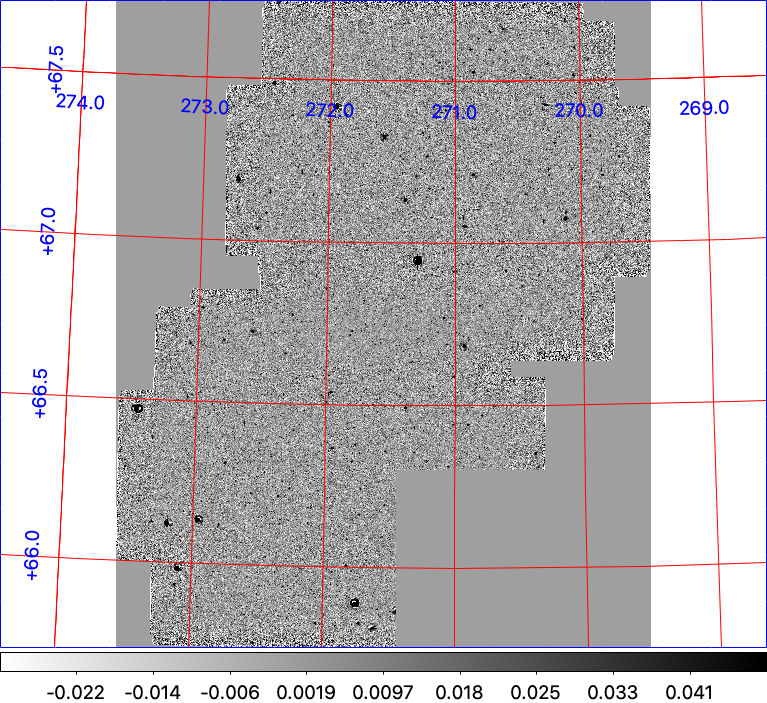}
	\includegraphics[width=\columnwidth]{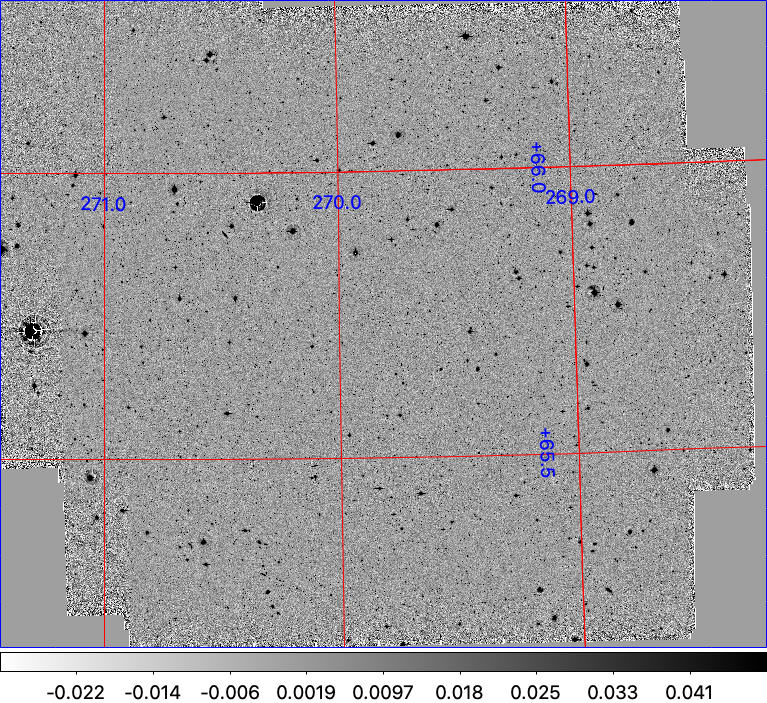}
	\includegraphics[width=\columnwidth]{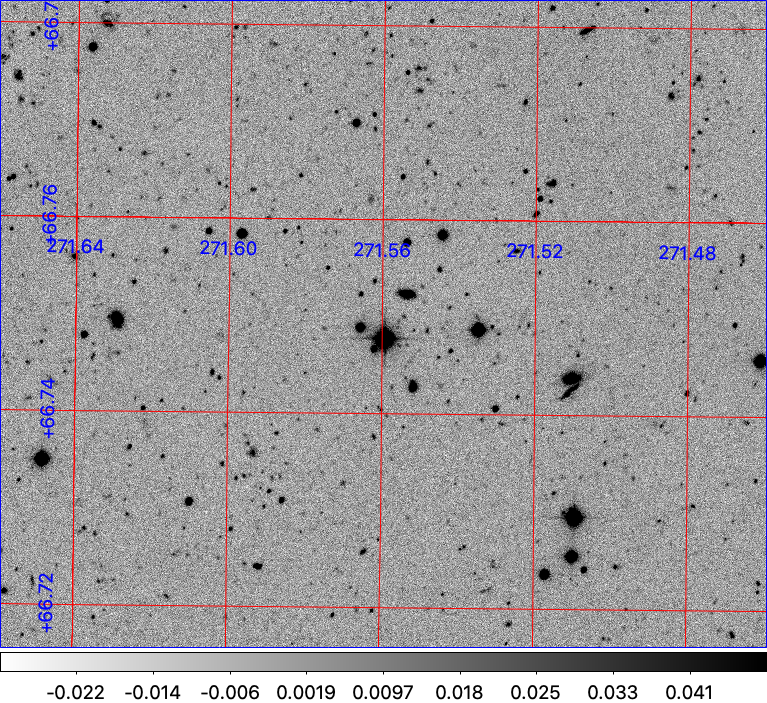}
	\includegraphics[width=\columnwidth]{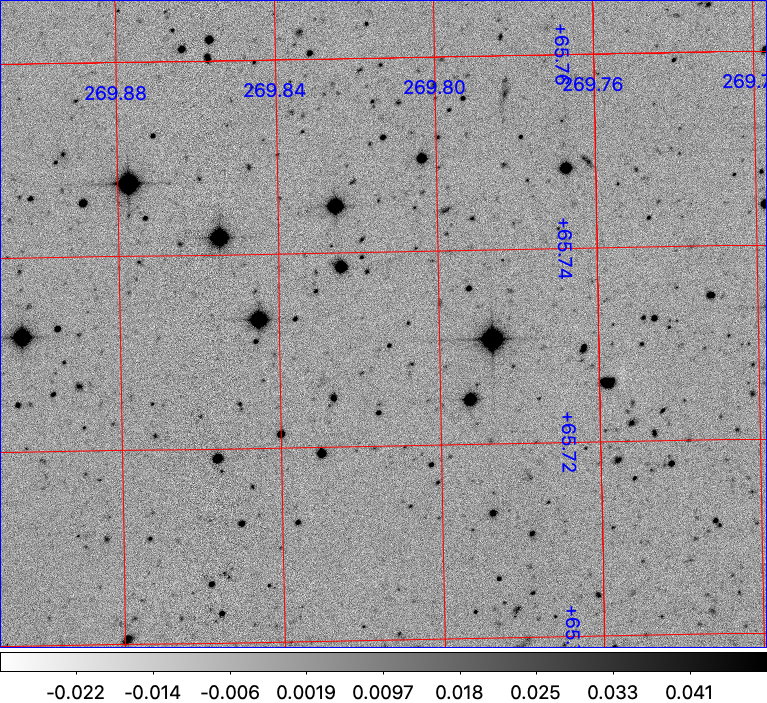}
    \caption{The cut co-added u-band images of the $AKARI$ NEP field. The images are cut into two panels based on the shallow (upper left) and the deep (upper right) observations. The lower left and the lower right panels show are the zoom-in images in the shallow and the deep regions, respectively. The coordinates are plotted in red grids with blue numbers. The images have the same scale of flux in the units of ADU sec$^{-1}$.}
    \label{fig:cut}
\end{figure*}

In this section, we describe the data quality of our $u$-band image and source extraction, by examining the 5$\sigma$ limiting magnitude, seeing, completeness, and source number counts. We use the 2-arcsec-diameter aperture magnitude only for the analysis of the limiting magnitude, while the magnitude used in other analyses and discussions is the Kron magnitude (i.e. MAG$\_$AUTO in \textsc{SExtractor}'s output). Since there is a significant difference of the integration time between two regions, i.e., 4,520 and 13,910 seconds (Section~\ref{obs}, Table~\ref{tab:obs}), from now on we separate the whole observed field into two regions (hereafter, the shallow and the deep) for the data quality analyses. The two regions are roughly cut by two lines, $\alpha$=18:05:48 and $\delta$=+66:17:56, and the master co-added image is cut into the shallow and the deep regions, shown in the upper left and the upper right panels in Fig.~\ref{fig:cut}, respectively. The lower left and lower right panels show the zoom-in images of the shallow and the deep images, respectively. The images are plotted in inverted colours so that faint sources might be more easily seen visually.

\subsection{Limiting magnitude}
\begin{figure*}
	\includegraphics[width=\columnwidth]{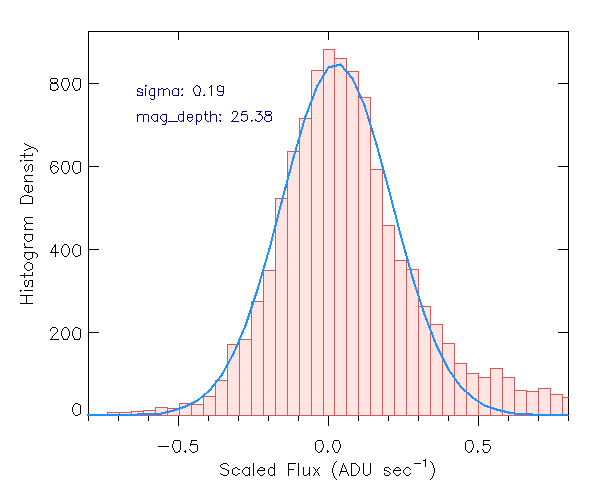}
	\includegraphics[width=\columnwidth]{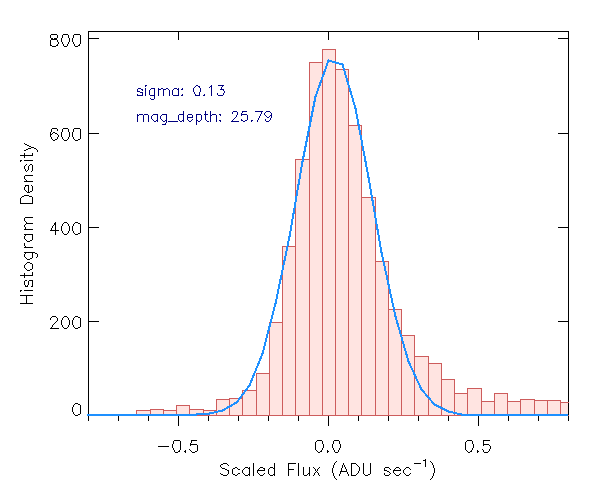}
    \caption{The Gaussian fits to the distributions of the randomly placed 2-arcsec-diameter aperture fluxes in the shallow (the left panel) and the deep (the right panel) regions. The flux in this figure is scaled by \textsc{Scamp} with the magnitude zero-point. The shallow region has the standard deviation of 0.19 and results in a 5$\sigma$ limiting magnitude of 25.38 mag. On the other hand, the standard deviation and the limiting magnitude in the deep region are 0.13 and 25.79 mag, respectively.}
    \label{fig:flux_dist}
\end{figure*}

We estimate the limiting magnitude ($MAG$; in AB system) by placing 10,000 2-arcsec\-diameter apertures  randomly in the co-added image. We masked out the pixels with value greater than 0.04 ADU sec$^{-1}$, which is corresponding to 28.78 mag/pixel (also see Fig.~\ref{fig:cut} for the pixel value), before placing the apertures to reduce the contamination from sources. Since the pixel value of the co-added image is calibrated with the zero-point, the sum of all the pixel values in an aperture is the scaled flux ($FLUX$), which means we can convert it to the magnitude with equation~\ref{eq:mag}. Thus, the 5$\sigma$ limiting magnitude is: 
\begin{equation}
    MAG_{\textrm{lim}} = 25.188 + 0.100 - 2.5\times \textrm{log}(5\sigma_{\textrm{f}}),
\end{equation}
where the 25.188 is the zero-point set in \textsc{SExtractor}, the 0.100 is the zero-point correction term obtained from the calibration with SDSS (Section~\ref{zpcorr}), and the $\sigma_{\textrm{f}}$ is the standard deviation obtained from the Gaussian fitting of the distribution of the 10,000 random aperture flux values.
Fig.~\ref{fig:flux_dist} shows the flux distributions from 20,000 random apertures in the shallow (the left panel) and the deep (the right panel) regions. The 5$\sigma$ detection limit of our observations are 25.38 and 25.79 mag in the shallow and the deep regions, respectively. The distributions have small tails at the bright end in both regions. We suspect that the tails exist because our bright pixel mask at 0.04 is not very strict. Nevertheless, using a stricter mask may result in an overestimation of limiting magnitude because the background becomes too clean (artificially suppressed). Since we do not want to overstate the depth of our image, we accept the result using the mask value 0.04, despite the existence of the tails in the distributions. Also, because the Gaussian distribution is fitting well at the faint end, we believe the tails do not affect the results of our limiting magnitude estimations too much.

We further provide an exercise to examine the limiting magnitude associated with the pixel mask of 0.04 ADU sec$^{-1}$. We consider the maximum contamination level that the aperture is exactly placed at the centre of a potential faint point source with the peak of 0.04 ADU sec$^{-1}$. We model the point spread function (PSF) of the point source by a two-dimensional Gaussian fitting discussed in the next section \ref{seeing}. Therefore, the total scaled flux estimated from the integral of two-dimensional Gaussian function is $0.04\times2\pi A_{2}A_{3}$, where the $A_{2}$, $A_{3}$ are the fitting parameters of the Gaussian function described in equation \ref{eq:gauss}. As the result, the scaled flux is 0.887 (or 1.159 in the deep region), which is corresponding to 25.42 (or 25.12) mag. The value is close to the limiting magnitude we obtained, and for most of the randomly placed apertures, their enclosed flux densities and magnitudes should be much fainter than this value.

\subsection{Seeing}
\label{seeing}
\begin{figure*}
	\includegraphics[width=\columnwidth]{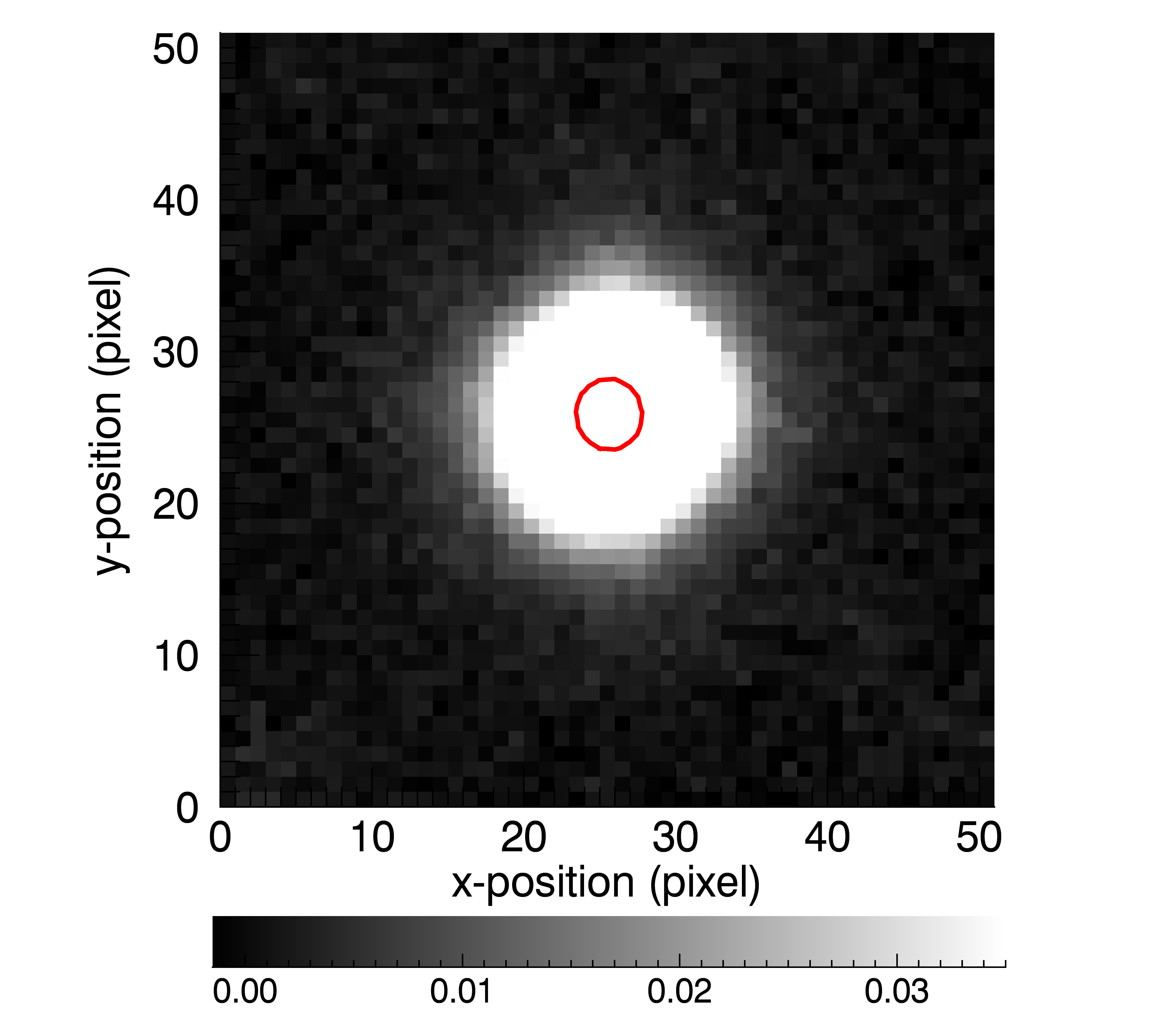}
	\includegraphics[width=\columnwidth]{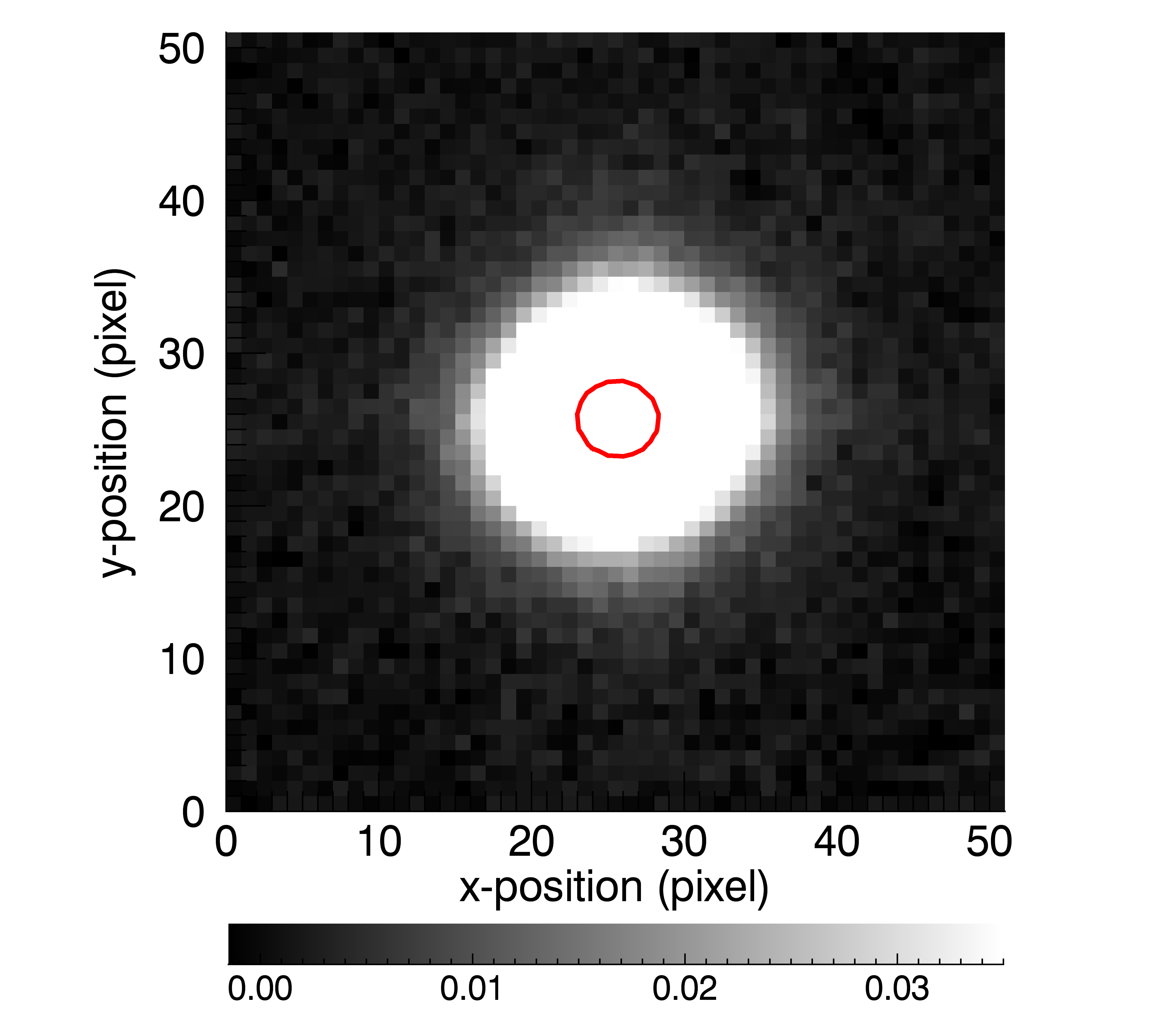}
    \caption{The images of the average-combined normalised point sources from the shallow (the left panel) and the deep (the right panel) regions. The size of each image is 51 pixels by 51 pixels. The pixel value of the image is scaled flux (in ADU sec$^{-1}$). The red ellipse is the contour at the half maximum.}
    \label{fig:PSF}
\end{figure*}

We randomly picked twenty normalised bright point sources, and created combined images by taking averages for the shallow and the deep regions. The definition of the brightness here is the range from 16 to 20 magnitudes, and a point source is defined if its CLASS$\_$STAR value, the output parameter from \textsc{SExtractor}, is larger than 0.9. All the sources are normalised to 20 magnitudes for making the average-combined image. The average-combined images of the bright point sources in the shallow and the deep regions are shown in Fig.~\ref{fig:PSF}. 

Astronomical seeing is usually estimated by the FWHM of the PSF. We used the two-dimensional Gaussian function as the approximation to the PSF. The average-combined point sources were fit with the two-dimensional Gaussian function $F(x,y)$ in the following equations.  
\begin{equation}
\label{eq:gauss}
\begin{split}
    F&(x,y) = A_{0} + A_{1}\textrm{e}^{\frac{-U}{2}}, \\
    U& = (\frac{x'}{A_{2}})^{2} + (\frac{y'}{A_{3}})^{2}, \\
    x'& = (x-A_{4})\textrm{cos}A_{6}-(y-A_{5})\textrm{sin}A_{6}, \\
    y'& = (x-A_{4})\textrm{sin}A_{6}+(y-A_{5})\textrm{cos}A_{6},
\end{split}
\end{equation}
where $x$ and $y$ are the position of pixel, and  $\{A_{1},A_{2},A_{3},A_{4},A_{5},A_{6} \}$ are the fitting parameters of the two-dimensional Gaussian function. The best-fitting results of the two average-combined point sources are shown in Table~\ref{tab:gauss}.    

\begin{table}
	\centering
	\caption{The best-fitting parameters of the two-dimensional Gaussian function with the average-combined point sources.}
	\label{tab:gauss}
	\begin{tabular}{lcr} 
		\hline
		Parameters & Shallow & Deep \\
		\hline
		$A_{0}$ & 0.007 & 0.007\\
		$A_{1}$ & 4.804 & 3.612\\
		$A_{2}$ & 1.812 & 2.231\\
		$A_{3}$ & 1.948 & 2.067\\
		$A_{4}$ & 25.62 & 25.65\\
		$A_{5}$ & 25.87 & 25.70\\
		$A_{6}$ & 0.000 & 0.000\\
		\hline
	\end{tabular}
\end{table}

The geometric mean value of $A_{2}$ and $A_{3}$ is used to derive the seeing FWHM (i.e. FWHM$=2\sqrt{2\textrm{ln}(2)\times A_{2}A_{3}}$). The seeing FWHMs in the shallow and the deep regions are 4.42 and 5.05 pixels (or 0.82 and 0.94 arcsec), respectively. The seeing FWHM in the deep region is broader possibly because there are more co-added frames (50 frames) in the deep region.

\subsection{Completeness: mock source extraction}
In this section we simply test the source extraction using the mock images combining the average-combined images created in Section \ref{seeing} as the sources with randomly picked backgrounds from the original co-added image. We made both the source image and the background image the same size, 51 pixels by 51 pixels, and stacked them to become the mock image of one source. We aligned all mock images of one source with different magnitudes and backgrounds into a big one. The sources were adjusted to different magnitudes ranging from 20 to 27 mag, while backgrounds were picked at 6 randomly chosen positions without any sources. Magnitude was set to vary in steps of 0.1, that is, there were 70 different values ranging from 20 to 27 mag. Thus with 6 backgrounds there were 420 combinations in total. To strengthen the statistical reliability, we enlarged the sample size 10 times, to 4200 sources. In order to judge the mock source extraction, we defined a detection is true if it satisfies the following criteria: \\
(1) The magnitude difference between the detected value and the true value is less than 0.2. \\
(2) The position difference between the detected value and the true value is less than 3 pixels.

\begin{figure}
	\includegraphics[width=\columnwidth]{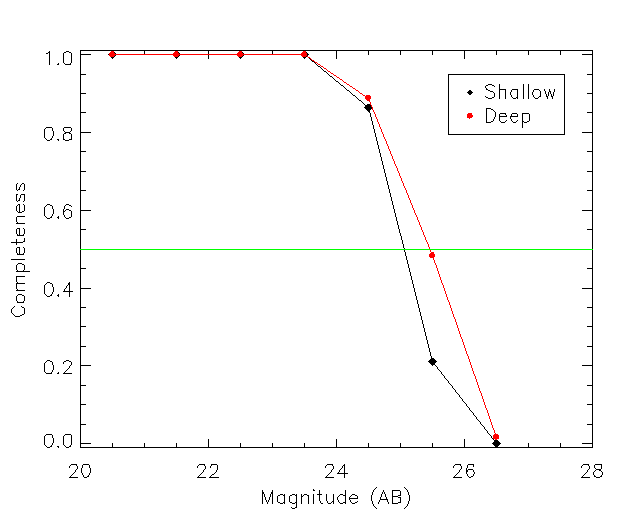}
    \caption{The completeness as functions of magnitude in the shallow (black diamonds) and the deep (red circles) regions. The interpolated magnitudes at the 50 per cent completeness level are 25.06 and 25.45 mag in the shallow and the deep regions, respectively.}
    \label{fig:completeness}
\end{figure}

The completeness, the success rate of mock source extraction, as a function of true magnitude is presented in Fig.~\ref{fig:completeness}. We use the same analysis procedure for the shallow and the deep regions, and the results are plotted in black squares and red circles. The bin size is 1.0 magnitude, and there are 600 sources in each bin. A light green horizontal line indicates the value of 0.5 (i.e. 50 per cent completeness). The interpolated values of 50 per cent completeness in the shallow and the deep regions are 25.06 and 25.45 mag, respectively.

\subsection{Source number counts}
\label{snc}
\begin{figure}
	\includegraphics[width=\columnwidth]{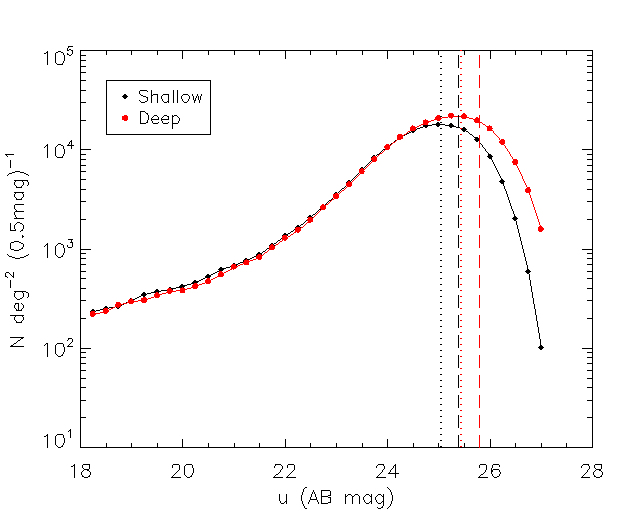}
    \caption{The source number counts of the $AKARI$ NEP u-band observation by CFHT MegaPrime/MegaCam. The observed region are separated into the shallow region (black diamonds) and the deep region (red circles) based on the total integration time. The 5$\sigma$ limiting magnitude and the 50 per cent completeness are labeled in the dash line and the dotted line, respectively.}
    \label{fig:SNC}
\end{figure}

In this section, we present the source number counts of the CFHT MegaPrime/MegaCam $u$-band observation in the $AKARI$ NEP field. The result is shown in Fig.~\ref{fig:SNC}, and as previous analyses, the sources number counts in the shallow (black diamond) and the deep (red circle) regions are discussed separately. The limiting magnitude and the 50 per cent completeness are overplotted in dotted and dashed lines. Note that since we adopted a relatively relaxed convolution filter for source extraction (see section \ref{se}), the number counts at the faint end may be overestimated. 

\section{Discussions}
\label{dis}
\subsection{Comparison with previous CFHT u*-band surveys}
\begin{figure*}
	\includegraphics[width=\columnwidth]{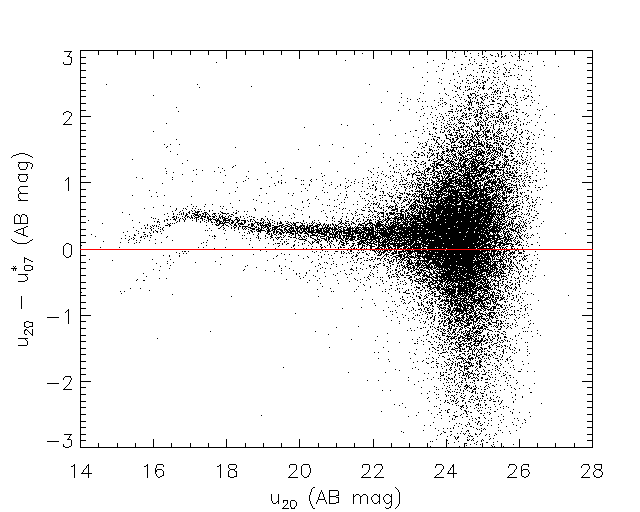}
	\includegraphics[width=\columnwidth]{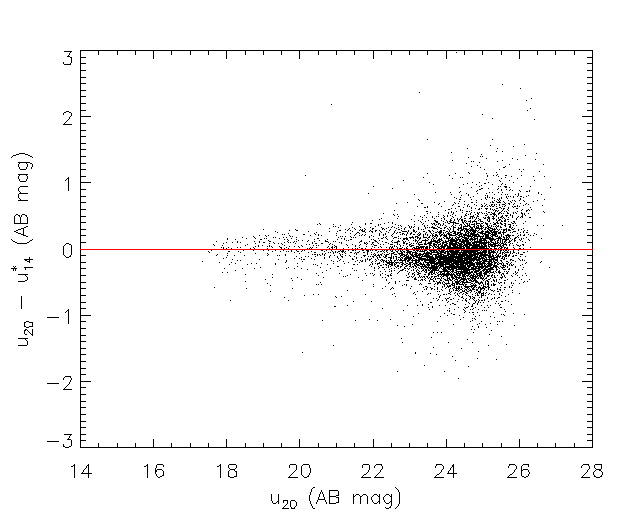}
    \caption{The magnitude comparisons of this work ($u_{20}$) with the previous $u^{*}$-band works, H07 ($u^{*}_{07}$) and O14 ($u^{*}_{14}$), in the AKARI NEP field. Every black dot represents a matched  source. The red horizontal line shows the positions that y-value equals to zero.}
    \label{fig:umatch}
\end{figure*}

\begin{figure*}
	\includegraphics[width=\columnwidth]{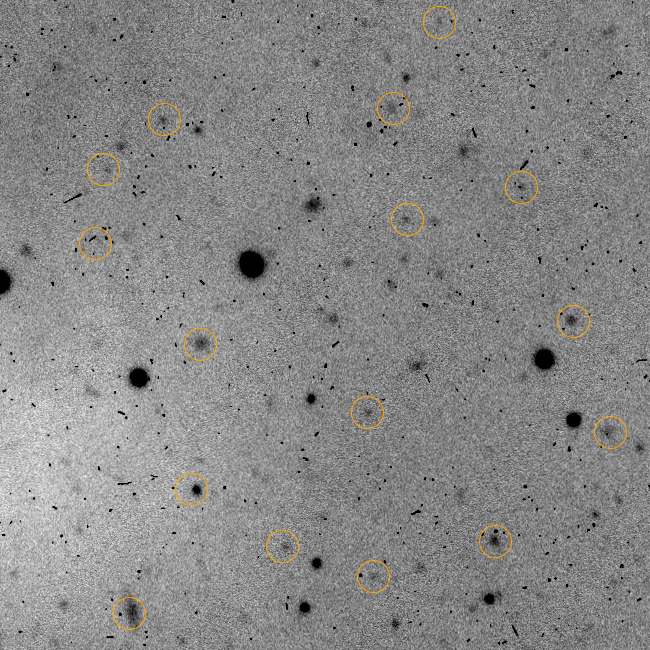}
	\includegraphics[width=\columnwidth]{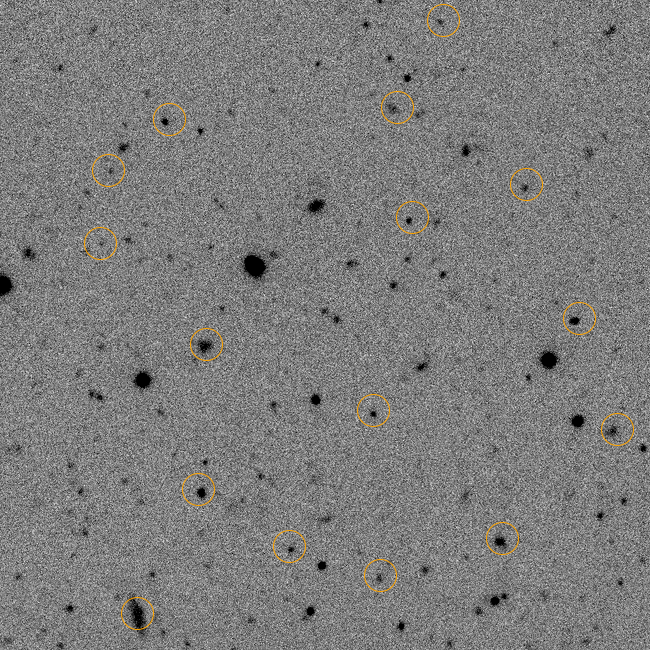}
    \caption{The example of the highly scattered sources ($|u_{20}-u^{*}_{07}|>1.0$) plotted in the image of H07 (left) and the image of this work (right). The images are plotted in inverted colours. The radius of the orange circle is 2 arcsec. The circles are placed according to the H07's coordinates.}
    \label{fig:u_ex}
\end{figure*}

Here we compare our result with other CFHT $u^{*}$-band catalogues from the previous observations in the $AKARI$ NEP field. First, we compare the measured magnitude of common sources. The sources from the previous works were matched with this work using a 1-arcsec radius, and 35,847 and 8,985 matches were produced between previous catalogues, H07 and O14, and this work, respectively. Fig.~\ref{fig:umatch} shows the magnitude comparison between this work ($u_{20}$) and the previous works, H07 ($u^{*}_{07}$) and O14 ($u^{*}_{14}$). A line with the y-value of 0 is plotted in red to indicate the magnitude consistency. The results show that there may be slight zero-point offsets between this work and the previous ones, but the offsets does not seem to be significant. However, there is a large scatter between $u_{20}$ and $u^{*}_{07}$ at faint magnitudes. We check the sources which are highly scattered ($|u_{20}-u^{*}_{07}|>1.0$) in the data images (Fig. \ref{fig:u_ex}). We are afraid that H07's limited observation and image quality account for the inconsistency. Our optimised dithering strategy, which performed observations of at least 16 frames, improves upon the observational strategy conducted by H07. Therefore, inevitably H07's image is more likely to be damaged by cosmic rays, bad pixels, or CCD edges. In the left panel (H07's image) of Fig.~\ref{fig:u_ex}, one can see that there are many artificial signals distributed around the image, which may possibly impact the source extraction. Moreover, many sources (orange circles) in H07's images are faint and vague or even not looking real. In conclusion, based on our improved observation and careful data reduction, we are confident that our data are more reliable.

\begin{figure}
	\includegraphics[width=\columnwidth]{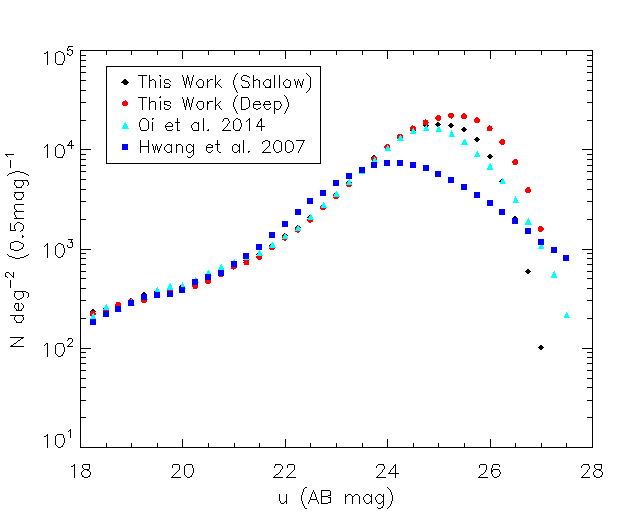}
    \caption{The source number counts of different CFHT MegaCam $u$-band/$u^{*}$-band surveys. The results from this work are plotted in black diamonds and red circles. The results from the previous surveys in the $AKARI$ NEP field, H07 and O14, are plotted in blue squares and green triangles, respectively.}
    \label{fig:SNC_compare}
\end{figure}

The source number counts comparison is shown in Fig.~\ref{fig:SNC_compare}. The results from this work are plotted in black diamonds and red circles as the same manner in Section~\ref{snc}. On the other hand, results from previous observations are plotted in green triangles (O14) and blue squares (H07). One can see that the number density in this work is higher than previous results at faint magnitudes, which implies that this work improves the depth of the u-band data in the $AKARI$ NEP field. This fact can be seen in terms of the limiting magnitude as well, which is the intrinsic depth from the image and independent to the source extraction. The limiting magnitude of the u-band data from O14 and H07 are 24.6 mag [5$\sigma$; 2 arcsec] and 25.9 mag [4$\sigma$; 1 arcsec], while this work reaches the limiting magnitude of 25.79 mag [5$\sigma$; 2 arcsec].

\section{Summary}
\label{sum}
This work, which is based on observations using the CFHT MegaPrime/MegaCam, provided the $u$-band source and HSC-based source catalogues in the $AKARI$ NEP field. The data were reduced in a standard manner, except that the convolution filter choice for the source extraction was relatively relaxed for the purpose of obtaining a better coordination with the deeper HSC sources. The data quality has been examined in the limiting magnitude, seeing, completeness, and source number counts. Compared with previous $u^{*}$-band surveys in the $AKARI$ NEP field, this work not only completed more of the previously unobserved area, but also achieved the deepest imaging. This promising improvement will aid future science studies in the NEP.

\section*{Acknowledgements}
We thank the anonymous reviewers for checking the paper carefully and providing insightful corrections and suggestions. T.-C. Huang thanks his colleague Ryan Lau for proofreading the manuscript. This work is based on observations obtained with MegaPrime/MegaCam, a joint project of CFHT and CEA/DAPNIA, at the Canada-France-Hawaii Telescope (CFHT) which is operated by the National Research Council (NRC) of Canada, the Institut National des Sciences de l'Univers of the Centre National de la Recherche Scientifique (CNRS) of France, and the University of Hawaii. The observations at the CFHT were performed with care and respect from the summit of Maunakea which is a significant cultural and historic site. T. Goto is supported by the grant 105-2122-M-007-003-MY3 and 108-2628-M-007-004-MY3 from Ministry of Science and Technology of Taiwan (MOST). T. Hashimoto is supported by the Centre for Informatics and Computation in Astronomy (CICA) at National Tsing Hua University (NTHU) through a grant from the Ministry of Education of the Republic of China (Taiwan).

\section*{Data availability}
The $u$-band source catalogue is available in Zenodo, at \url{https://dx.doi.org/10.5281/zenodo.3980635}. Other data underlying this article will be shared on reasonable request to the corresponding author.







\appendix

\section{Filter Choice for Source Extraction}
\label{apdx}
In this section, we discuss the filter file choice in \textsc{SExtractor} for the source extraction. As mentioned in Section~\ref{se}, we intend to detect more faint sources because this $u$-band data is going to be used along with the deeper optical data observed by HSC. Here we compare the source extractions using the filter Gauss2.0$\_$3x3 and Gauss2.0$\_$5x5 by checking the match with HSC. 

\begin{figure}
	\includegraphics[width=\columnwidth]{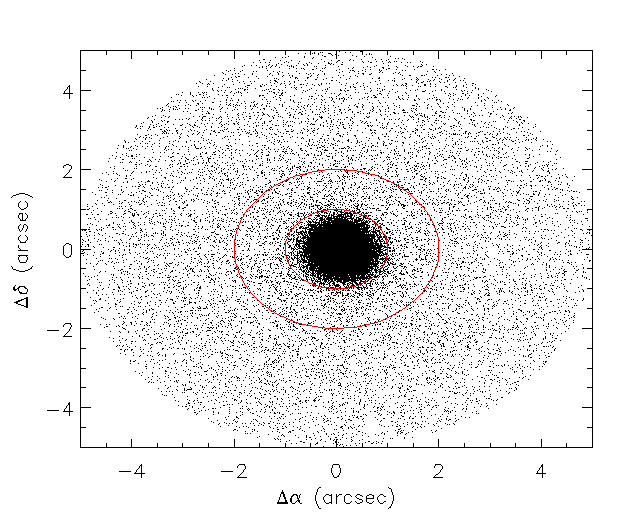}
    \caption{The coordinates difference between the u-band coordinates (Gauss2.0$\_$3x3) and the HSC coordinates. The inner and the outer red circles plot the 1-arcsec and 2-arcsec radius, respectively.}
    \label{fig:pos}
\end{figure}

\begin{figure*}
	\includegraphics[width=\columnwidth]{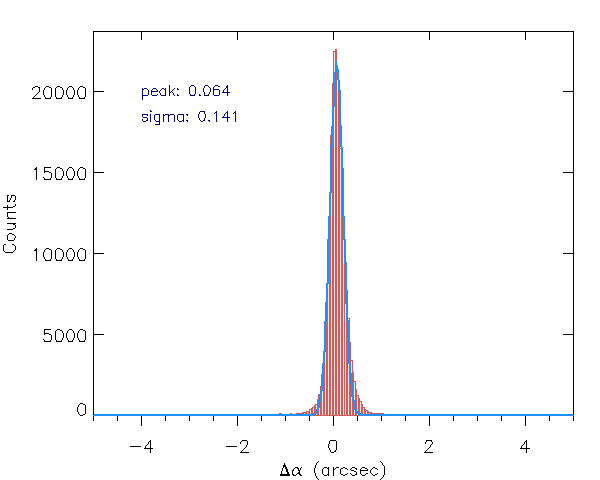}
	\includegraphics[width=\columnwidth]{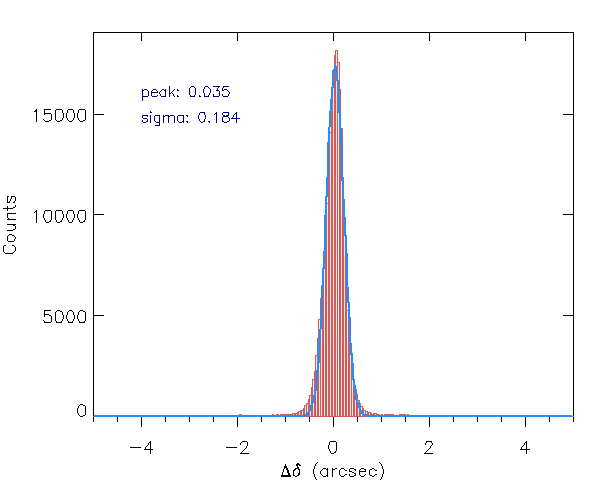}
    \caption{The Gaussian fittings with the distributions of the $\alpha$ (the left panel) and the $\delta$ difference (the right panel) in the Fig.~\ref{fig:pos}. $\alpha$ has a peak at 0.064 and a standard deviation of 0.141, while $\delta$ peaks at 0.035 and deviates as 0.184.}
    \label{fig:fit}
\end{figure*}

\begin{figure*}
	\includegraphics[width=0.48\columnwidth]{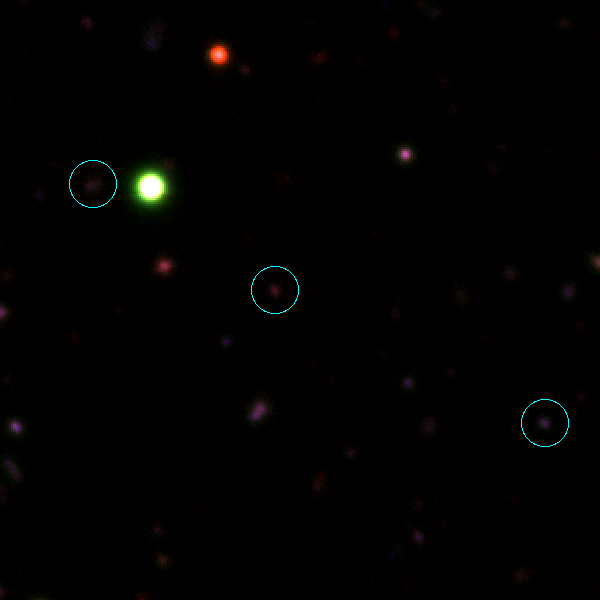}
	\includegraphics[width=0.48\columnwidth]{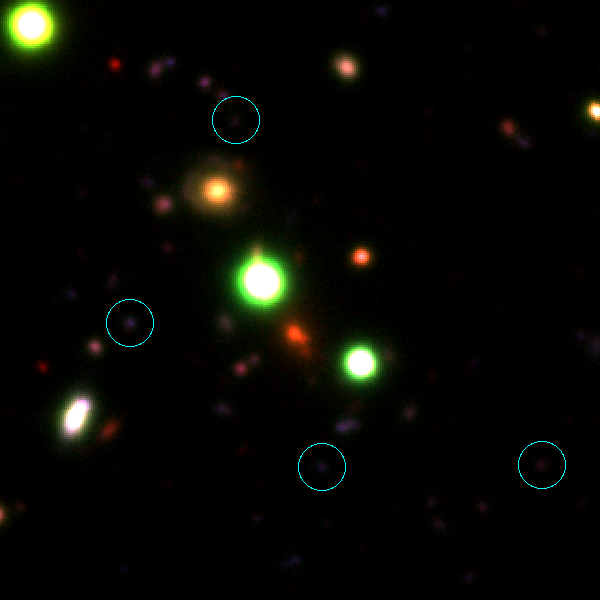}
	\includegraphics[width=0.48\columnwidth]{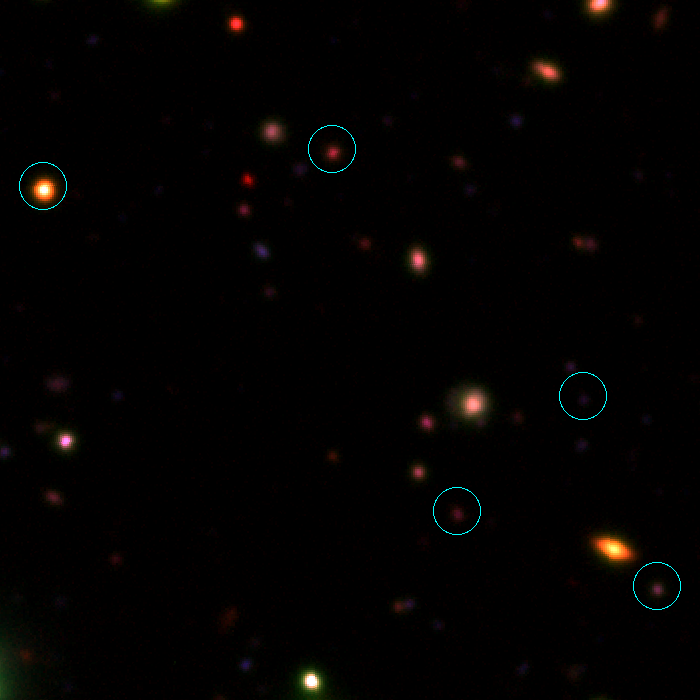}
	\includegraphics[width=0.48\columnwidth]{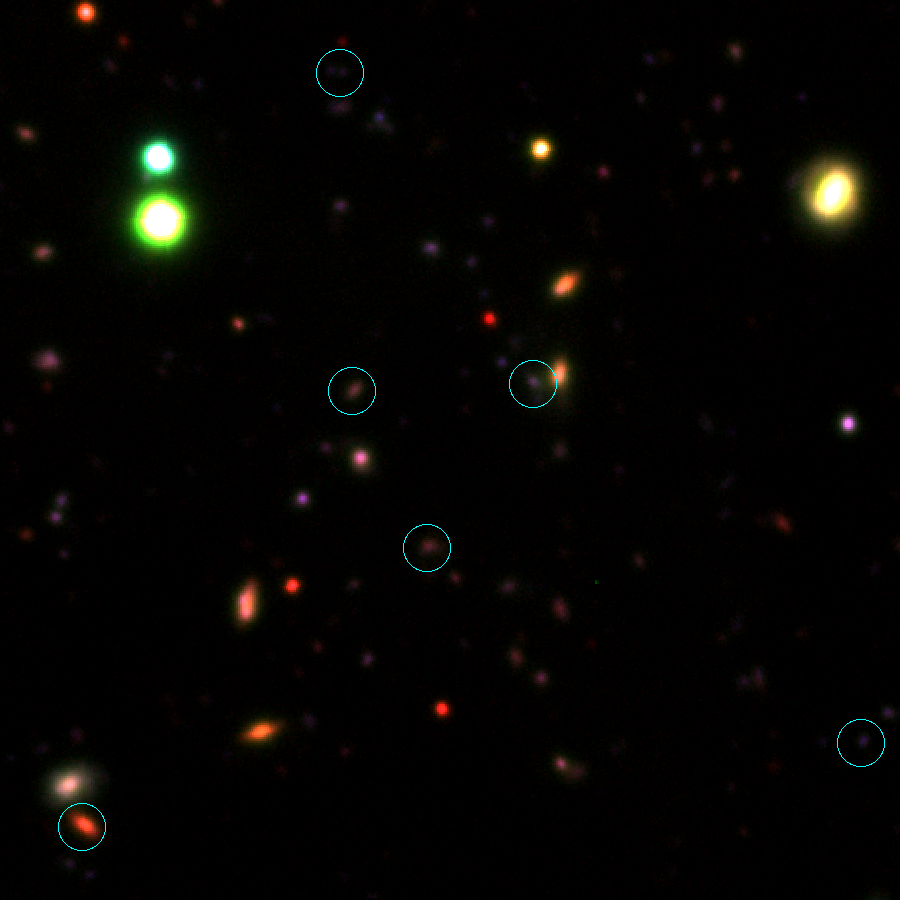}
    \caption{The example sources (encircled by cyan circles) that are missed in the source extraction using the Gauss2.0$\_$5x5 filter but detected using Gauss2.0$\_$3x3. The colour image is generated by three colours, blue ($g$ band), green ($r$ band), and red ($i$ band). The radius of the cyan circle is 2 arcsec.}
    \label{fig:srcs}
\end{figure*}

A correction to the astrometric difference of the $u$-band catalogue and the HSC catalogue is preferable, so we first matched the sources of the 2 catalogues in a 5" radius (Fig.~\ref{fig:pos}). From the match we obtained the distributions of $\alpha$ and $\delta$ differences, and the Gaussian fittings were then applied to them to obtain values of the systematic positional offset (Fig.~\ref{fig:fit}). We further matched the sources again using a 1-arcsec radius to obtain the final match catalogue. Table~\ref{tab:match} summarises the numbers of the CFHT u-band sources and the sources matched with HSC using the two different filters for source extraction. Using Gauss2.0$\_$5x5 and Gauss2.0$\_$3x3 produce 319,965 and 351,635 sources, and 236,024 and 250,468 sources have a match with HSC, respectively. In other words, using the filter Gauss2.0$\_$3x3 increases the number of sources by 31,670, and almost half of them, 14,444 sources, have a match with HSC, which means it is highly possible that they are real sources but not fake detection.
To confirm the result of the matched number, we visually checked the detected sources in the RGB colour image stacked by HSC g-, r-, and i-band images as blue, green, and red, respectively. Fig.~\ref{fig:srcs} shows some examples of the sources detected by using the filter Gauss2.0$\_$3x3 but missed by using Gauss2.0$\_$5x5. Most of the sources seem to be real sources in the HSC image, so it is likely that using Gauss2.0$\_$3x3 can really detected the faint sources as we expected. 
Consequently, we accept the trade-off that increasing the number of sources might also increases the number of fake detection at the same time. Therefore, we caution that the user of this u-band catalogue should be careful about the usage of the faint sources. 

\begin{table}
	\centering
	\caption{The filter comparison of the numbers of the sources detected in the $u$-band and the sources matched with HSC.}
	\label{tab:match}
	\begin{tabular}{lcr} 
		\hline
		Convolution Filter & $\#$ of sources & $\#$ of Matches with HSC \\
		\hline
		Gauss2.0$\_$5x5 & 319,965 & 238,755\\
		Gauss2.0$\_$3x3 & 351,635 & 253,670\\
		Difference & 31,670 & 14,915\\
		\hline
	\end{tabular}
\end{table}

\section{Catalogues and Formats}
\label{cat}
We provide two source catalogues, a $u$-band source catalogue and an HSC-based source catalogue, for potential users who may have an interest in studying extragalactic astronomy in the AKARI NEP field. The $u$-band source catalogue contains 351,635 sources, and 290,373 of them are not flagged. Table~\ref{tab:ex} shows the example rows of the $u$-band source catalogue, and it columns are briefly described in the following. \\
Column (1): identification number. \\
Columns (2) and (3): right ascension and declination (in degree). \\
Columns (4) and (5): magnitude and errors (in AB system). \\
Column (6): CLASS$\_$STAR, which is a star classifier based on the morphology of a source.\\
Column (7): source extraction flag, a 8-bit integer indicating that if a source is extracted with warnings. \\
The HSC-based source catalogue has the same first seven columns as the $u$-band catalogue. The description of the rest columns are listed below.
Column (8) and (9): right ascension and declination (in degree) from HSC detection. \\
Column (10) and (11): HSC $g$-band magnitude and errors (in AB system). \\
Column (12) and (13): HSC $r$-band magnitude and errors (in AB system).\\
Column (14) and (15): HSC $i$-band magnitude and errors (in AB system).\\
Column (16) and (17): HSC $z$-band magnitude and errors (in AB system).\\
Column (18) and (19): HSC $Y$-band magnitude and errors (in AB system).\\
Table~\ref{tab:HSC} shows a few rows of the HSC source catalogue as an example. Note that the digits of coordinates and magnitudes are truncated in this table in order to make the catalogue fit in one page. 
\begin{table*}
	\centering
	\caption{A few rows of the $u$-band source catalogue. The columns CS and f, represent the CLASS$\_$STAR value and flag, respectively.}
	\label{tab:ex}
	\begin{tabular}{lcccccr} 
		\hline
		ID & $\alpha$ & $\delta$ & $u$  & $u$ err. & CS & f \\
		\hline
		1 & 269.7556402 & +65.1727241 & 14.1269 & 0.0001 & 0.812 & 3\\
		2 & 269.0409577 & +65.1743737 & 14.1432 & 0.0001 & 0.805 & 3\\
		3 & 269.9828576 & +65.1820611 & 15.3149 & 0.0001 & 0.239 & 0\\
		4 & 270.8800236 & +65.1686552 & 23.4442 & 0.0431 & 0.996 & 1\\
		5 & 270.7984152 & +65.1689814 & 23.7094 & 0.0578 & 0.645 & 1\\
		6 & 270.6406487 & +65.1719293 & 24.4598 & 0.1844 & 0.481 & 2\\
		7 & 270.6994843 & +65.1724827 & 23.3358 & 0.1210 & 0.636 & 2\\
		8 & 270.7547065 & +65.1701884 & 23.4640 & 0.0481 & 0.983 & 0\\
		9 & 270.6491622 & +65.1705864 & 22.4449 & 0.0182 & 0.996 & 0\\
		10 & 270.7939588 & +65.1713746 & 24.2137 & 0.1285 & 0.439 & 0\\
		\hline
	\end{tabular}
\end{table*}

\begin{table*}
	\centering
	\caption{A few rows of the HSC-based source catalogue. The columns CS and f, represent the CLASS$\_$STAR value and flag, respectively.}
	\label{tab:HSC}
	\rotatebox{90}{
	\begin{tabular}{lcccccccccccccccccr} 
		\hline
		ID & $\alpha$ & $\delta$ & $u$ & $u$ err. & CS & f & HSC $\alpha$ & HSC $\delta$ &$g$ & $g$ err. & $r$ & $r$ err. & $i$ & $i$ err. & $z$ & $z$ err. & $Y$ & $Y$ err.\\
		244631&273.1532&+66.72558&24.1607&0.1010&0.648&0&273.1532&66.72556&23.6885&0.0139&23.4685&0.0120&23.5250&0.0852&23.1761&0.1129&22.8928&0.1293\\
244632&272.9458&+66.72837&24.4496&0.1207&0.101&0&272.9457&66.72836&24.0013&0.0118&23.1344&0.0088&22.9475&0.0191&22.7994&0.0331&22.7437&0.0578\\
244633&271.4180&+66.74085&24.8887&0.1127&0.488&0&271.4180&66.74079&24.2895&0.0079&24.1573&0.0156&23.9494&0.0281&23.3736&0.0374&23.1003&0.0483\\
244634&272.6394&+66.73219&24.1340&0.0526&0.581&0&272.6394&66.73219&23.8931&0.0047&23.8459&0.0125&23.7877&0.0215&23.7016&0.0385&23.3397&0.0450\\
244635&271.9581&+66.73840&25.6157&0.1556&0.351&0&0&0&0&0&0&0&0&0&0&0&0&0\\
244636&271.9432&+66.73877&26.0340&0.1859&0.348&0&271.9432&66.73874&24.7200&0.0182&24.2725&0.0290&23.5742&0.0339&23.2007&0.0403&22.6848&0.0387\\
244637&272.0030&+66.73823&25.9995&0.1370&0.405&0&272.0030&66.73825&25.0234&0.0150&24.7528&0.0315&24.5834&0.0413&24.3590&0.0794&23.6669&0.0767\\
244638&271.0122&+66.74067&22.8342&0.0212&0.522&0&271.0122&66.74069&22.5616&0.0016&22.2210&0.0030&21.7324&0.0035&21.4215&0.0058&21.6000&0.0106\\
244639&272.8397&+66.72987&23.8287&0.0705&0.530&0&0&0&0&0&0&0&0&0&0&0&0&0\\
244640&270.7374&+66.74144&25.2457&0.1211&0.453&0&270.7374&66.74145&24.7036&0.0084&24.2702&0.0183&23.8975&0.0216&23.4734&0.0308&23.2898&0.0403\\
		\hline
	\end{tabular}
	}
\end{table*}


\bsp	
\label{lastpage}
\end{document}